\input{aipcheck}

\documentclass[
    ,final            
  ]
  {aipproc}

\layoutstyle{6x9}

\begin{document}

\title{GRB Afterglows in the Deep Newtonian Phase}

\author{Y. F. Huang}{
  address={Department of Astronomy, Nanjing University, Nanjing 210093, China}
}

\author{K. S. Cheng}{
  address={Department of Physics, the University of Hong Kong, Hong Kong, China}
}

\author{Z. G. Dai}{
  address={Department of Astronomy, Nanjing University, Nanjing 210093, China}
}

\author{T. Lu}{
  address={Purple Mountain Observatory, Chinese Academy of Sciences, Nanjing 210008, China}
}

\begin{abstract}
 In many GRBs, afterglows have been observed for months 
or even years. It deserves noting that at such late stages, the 
remnants should have entered the deep Newtonian phase, during 
which the majority of shock-accelerated electrons will no longer 
be highly relativistic. However, a small portion of electrons are 
still ultra-relativistic and capable of 
emitting synchrotron radiation. Under the assumption that the 
electrons obey a power-law distribution according to their kinetic 
energy (not simply the Lorentz factor), we calculate optical 
afterglows from both isotropic fireballs and beamed ejecta, 
paying special attention to the late stages. In the beamed cases, 
it is found that the light curves are universally characterized 
by a flattening during the deep Newtonian phase. Implication of 
our results on orphan afterglows is also addressed. 
\end{abstract}

\maketitle


\section{Importance of Newtonian phase}

GRBs have been recognized as the most relativistic phenomena in the Universe.
In 1997, Wijers et al. (1997) once 
discussed GRB afterglows of the non-relativistic phase. However, for quite 
a long period, many authors were obviously beclouded by the powerfulness 
of GRBs and emission in the non-relativistic phase was generally omitted. 
In 1998, Huang et al. (1998) stressed the importance of Newtonian phase for
the first time. In fact, the Lorentz factor of GRB blastwave evolves as 
\begin{equation}
\label{eq1}
   \gamma \approx (200 - 400) E_{51}^{1/8} n_0^{-1/8} t_{\rm s}^{-3/8}, 
\end{equation}
in the ultra-relativistic phase. It is clear that the shock will enter the 
trans-relativistic phase within several months, and will become 
non-relativistic soon after that. Fig. 1a illustrates the 
condition clearly. Today, this point has been realized by more and more 
authors(Livio \& Waxman 2000; Frail et al. 2000; Dermer et al. 2000;
Dermer \& Humi 2001; Piro et al. 2001; in't Zand et al. 2001; 
Panaitescu \& Kumar 2003; Zhang \& M\'esz\'aros 2003). 

\begin{figure}
  \includegraphics[height=.35\textheight,width=0.28\textheight,angle=-90]{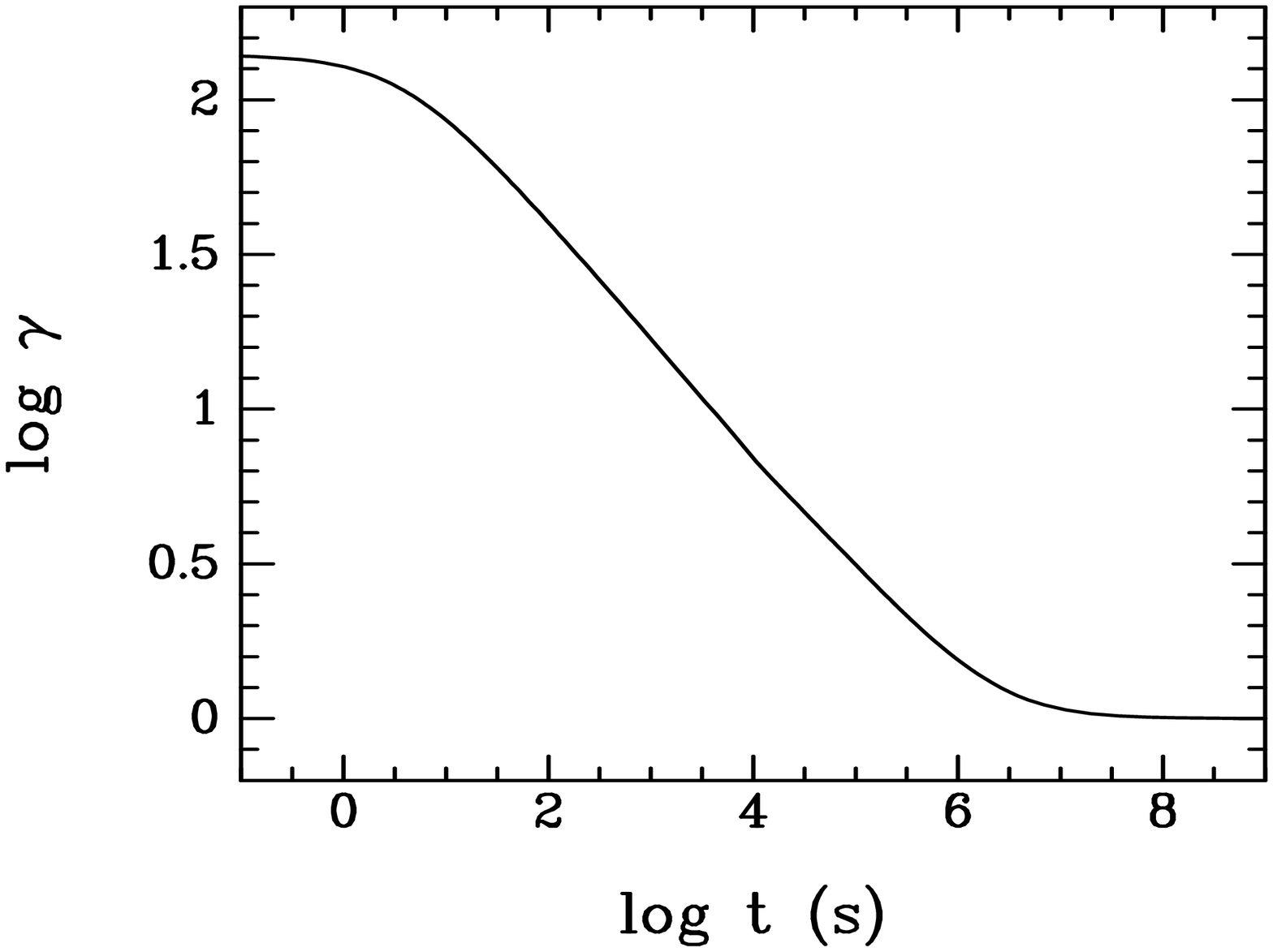}
  \includegraphics[height=.37\textheight,width=0.28\textheight,angle=-90]{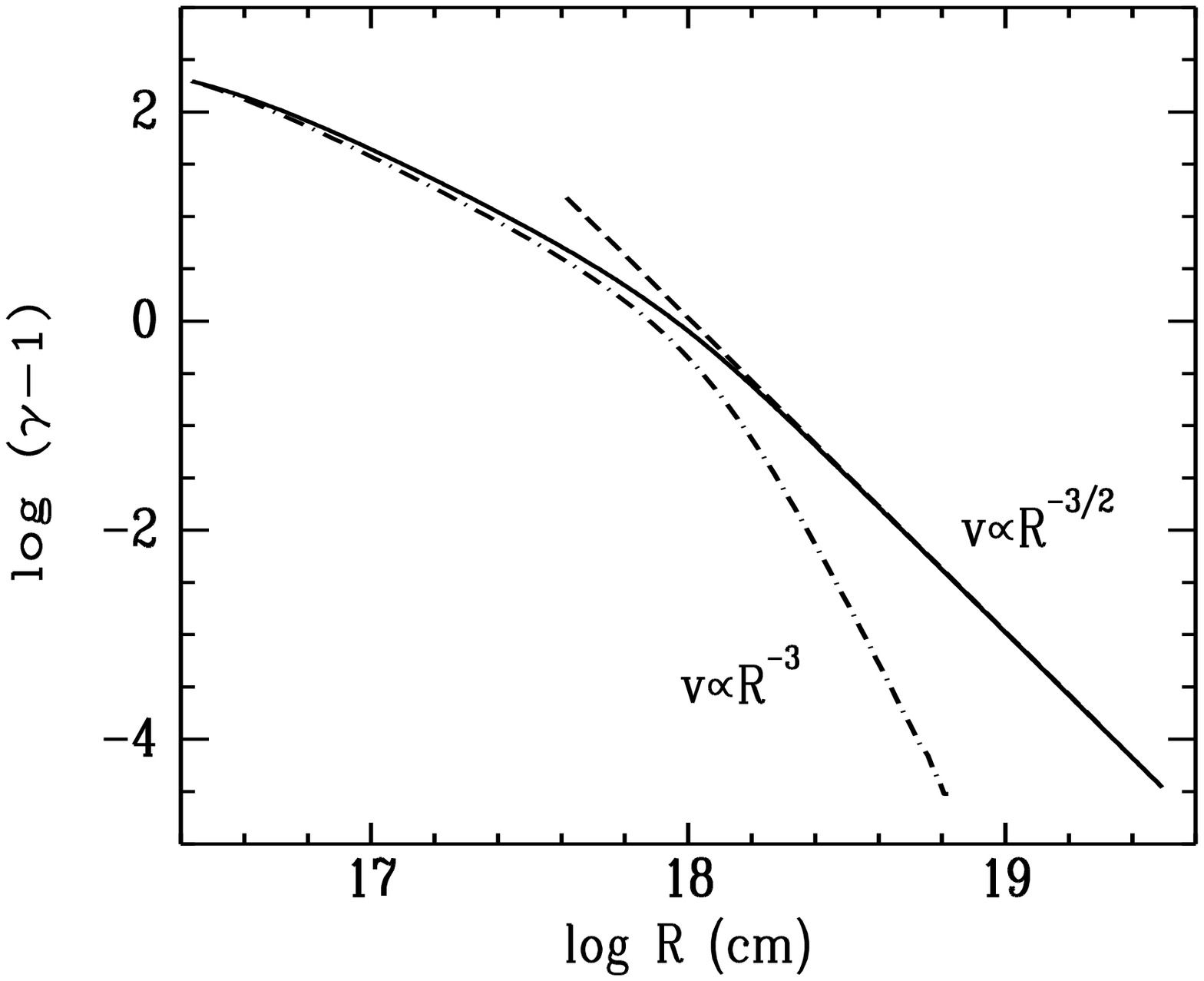}
  \caption{{\bf(a)} Left panel, evolution of an adiabatic fireball
  ($E_0=10^{51}$ ergs, and $n=1$ cm$^{-3}$, Huang et al. 1998), 
    which becomes non-relativistic in a few months; {\bf (b)} Right panel, 
 Eq. 2 (solid line) is correct in both the relativistic and 
 the Newtonian phases (also see Huang 2000).  }
\end{figure}

\section{Model}

A refined generic dynamical model has been proposed by Huang et al. (1999), which
is mainly characterized by
\begin{equation}
\label{eq2}
\frac{d \gamma}{d m} = - \frac{\gamma^2 - 1}
       {M_{\rm ej} + \epsilon m + 2 ( 1 - \epsilon) \gamma m}.
\end{equation}
Fig. 1b shows clearly that this equation is applicable in both the 
ultra-relativistic phase and the Newtonian phase. For a realistic 
description of the overall dynamical evolution of isotropic fireballs and 
collimated jets, we refer to Huang et al. (1999, 2000a, b).

GRB afterglows mainly come from synchrotron emission of shock-accelerated 
electrons. In the ultra-relativistic case, these electrons are 
generally assumed to distribute as 
$d N_{\rm e}'/d \gamma_{\rm e} \propto \gamma_{\rm e}^{-p}$, 
with $\gamma_{\rm e,min} \sim \xi_{\rm e} (\gamma-1) m_{\rm p}/m_{\rm e}$.
However, we noticed that $\gamma_{\rm e, min}$ will typically
be less than 2.0 when $t \geq $ a few months. This means most electrons 
will no longer be ultra-relativistic in the deep Newtonian phase 
(Huang \& Cheng 2003). 

We have suggested that the correct distribution 
function that is also applicable in the deep Newtonian phase should be 
(Huang \& Cheng 2003), 
\begin{equation}
\label{eq10}
\frac{d N_{\rm e}'}{d \gamma_{\rm e}} \propto (\gamma_{\rm e} - 1)^{-p} ,
\;\;\; (\gamma_{\rm e,min} \leq \gamma_{\rm e} \leq \gamma_{\rm e,max}).
\end{equation}
In the deep Newtonian phase, most electrons are 
now non-relativistic and their cyclotron radiation cannot be observed 
in the optical bands. But there are still many relativistic electrons
capable of emitting synchrotron radiation. With the help of Eq. (\ref{eq10}), 
optical afterglows can be calculated conveniently by integrating 
synchrotron emission from those electrons with Lorentz factors above 
a critical value ($\gamma_{\rm e,syn}$, Huang \& Cheng 2003). 

\section{Numerical results}

\begin{figure}
  \includegraphics[height=.26\textheight]{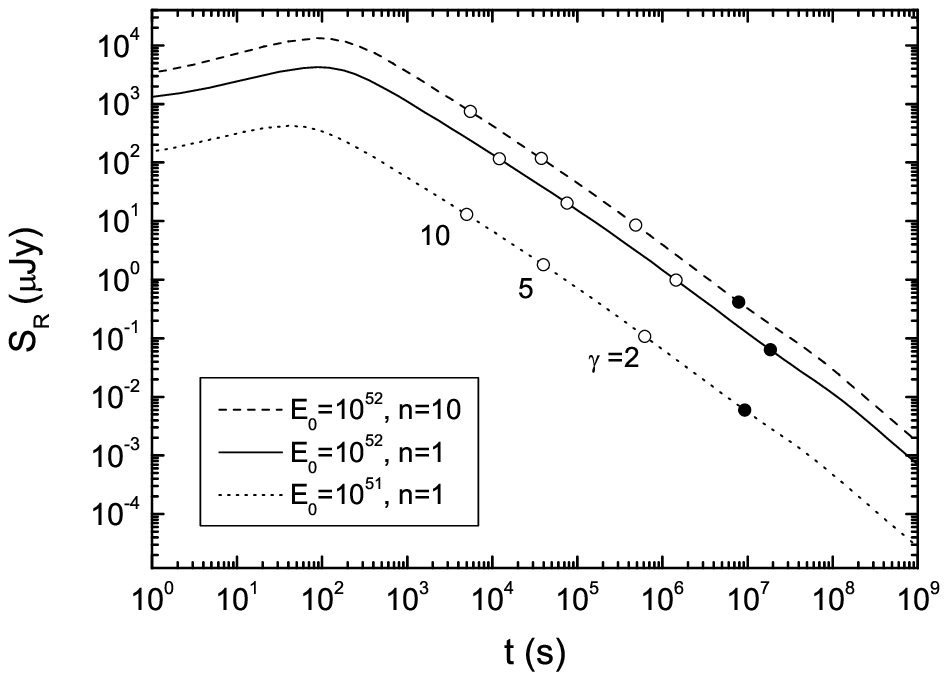}
  \includegraphics[height=.26\textheight]{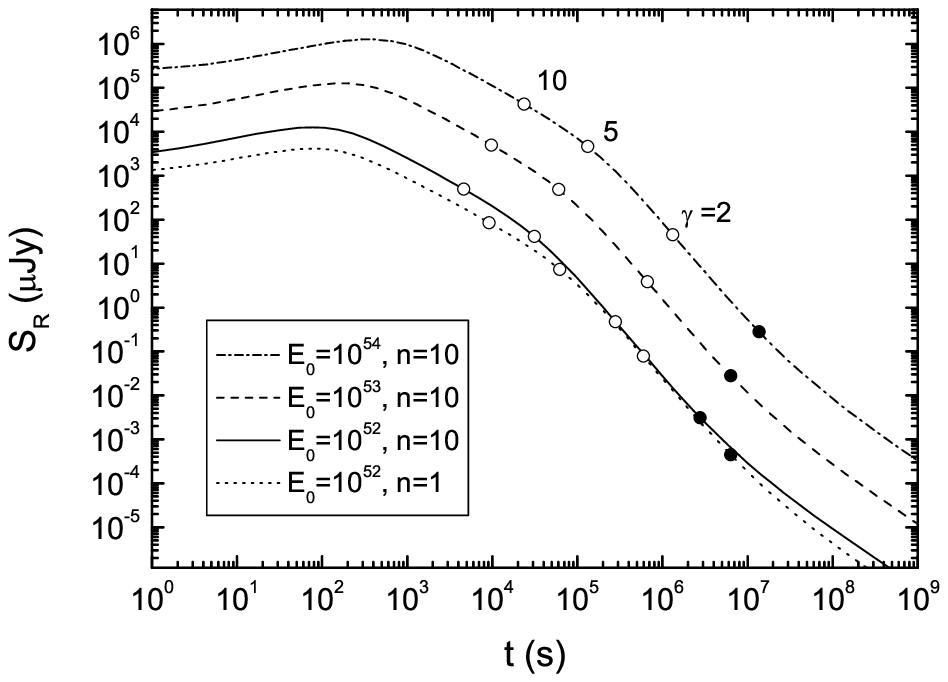}
  \caption{R-band optical afterglows from isotropic fireballs 
({\bf a}, left panel) and conical jets ({\bf b}, right panel) (Huang \& Cheng 2003). 
Black dot on each light curve indicates the moment when 
$\gamma_{\rm e,min} = \gamma_{\rm e,syn} \equiv 5$, and 
open circles mark the time when the bulk Lorentz 
factor $\gamma = 2,$ 5 and 10 respectively.}
\end{figure}

We present our numerical results in Fig 2.
Note that the light curves in Fig. 2a steepens slightly in the deep Newtonian 
phase. It is consistent with the analytical solution of, 
\begin{equation}
\label{13}
  S_{\rm R} \propto \left \{
   \begin{array}{ll}
 t^{(3-3p)/4}\,, \,\,\,\, & (\gamma \gg 1), \\
 t^{(21-15p)/10}\,, \,\,\,\, & (\beta \ll 1).
   \end{array}
   \right. 
\end{equation}
The light curve in Fig. 2b flattens in the deep Newtonian phase, 
which is also consistent with the analytical solution (Livio \& Waxman 2000),
\begin{equation}
\label{12}
  S_{\rm R} \propto \left \{
   \begin{array}{ll}
 t^{(3-3p)/4}\,, \,\,\,\, & (\gamma > 1/ \theta_0), \\
 t^{-p}\,, \,\,\,\, & (\gamma < 1/\theta_0 \; {\rm and} \; \beta \sim 1), \\
 t^{(21-15p)/10}\,, \,\,\,\, & (\beta \ll 1).
   \end{array}
   \right.
\end{equation}

\section{Implications on orphan afterglows}

Orphan afterglows are regarded as a useful tool for measuring the beaming angle
of GRBs (Rhoads 1997; Perna \& Loeb 1998; M\'esz\'aros et al. 1999; Grindlay 1999;
Lamb 2000; Totani \& Panaitescu 2002; Levinson et al. 2002; Nakar et al. 2002; 
Granot et al. 2002; Rhoads 2003; Yamazaki et al. 2003).
However, Huang et al. (2002) pointed out that there may exist large numbers of 
failed GRBs, i.e., fireballs with initial Lorentz factors $1 \ll \gamma_0 \ll 
100$, which fail to produce GRBs but are likely to give birth to X-ray flashes. 
The simple discovery of orphan afterglows then does not necessarily mean that
GRBs are highly collimated. 

To judge whether an orphan afterglow comes from a failed GRB or a jetted but 
off-axis GRB, a $\log S_{\rm R}$ --- $\log t$ light curve will be helpful. 
However, such a log-log light curve is usually not available for orphan afterglows, 
since the trigger time is unknown. Fig. 3 illustrates the 
effect of the uncertainty of the trigger time on the light curve. 

To overcome the problem, Huang et al. (2002) suggested that the most important thing
is to monitor the orphan for a relatively long period. Obviously, 
the calculation of afterglows in the deep Newtonian phase is necessary in the studies 
of orphan afterglows. 

\begin{figure}
\centerline{
  \includegraphics[height=.37\textheight,width=0.26\textheight,angle=-90]{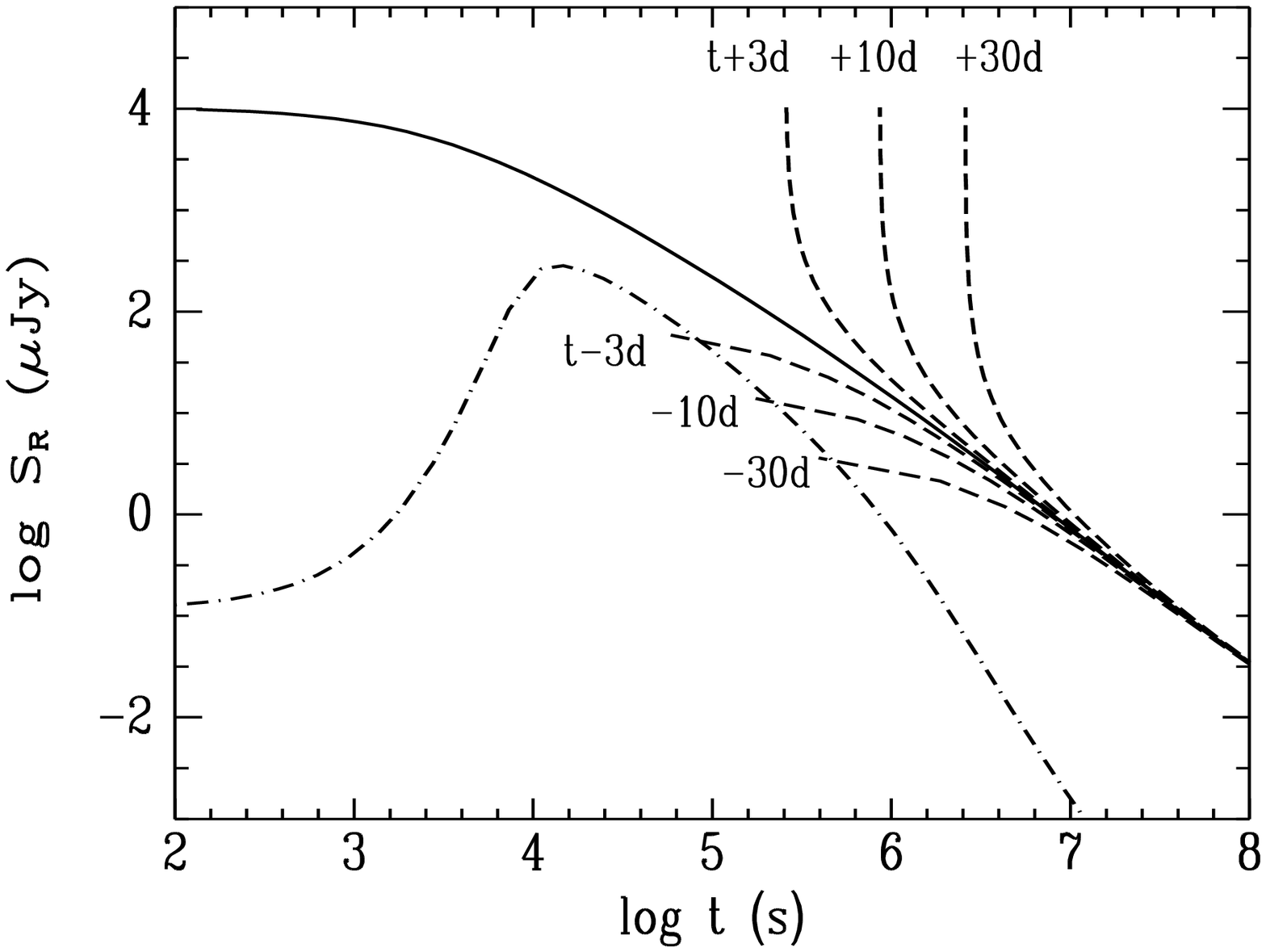}
  \includegraphics[height=.37\textheight,width=0.26\textheight,angle=-90]{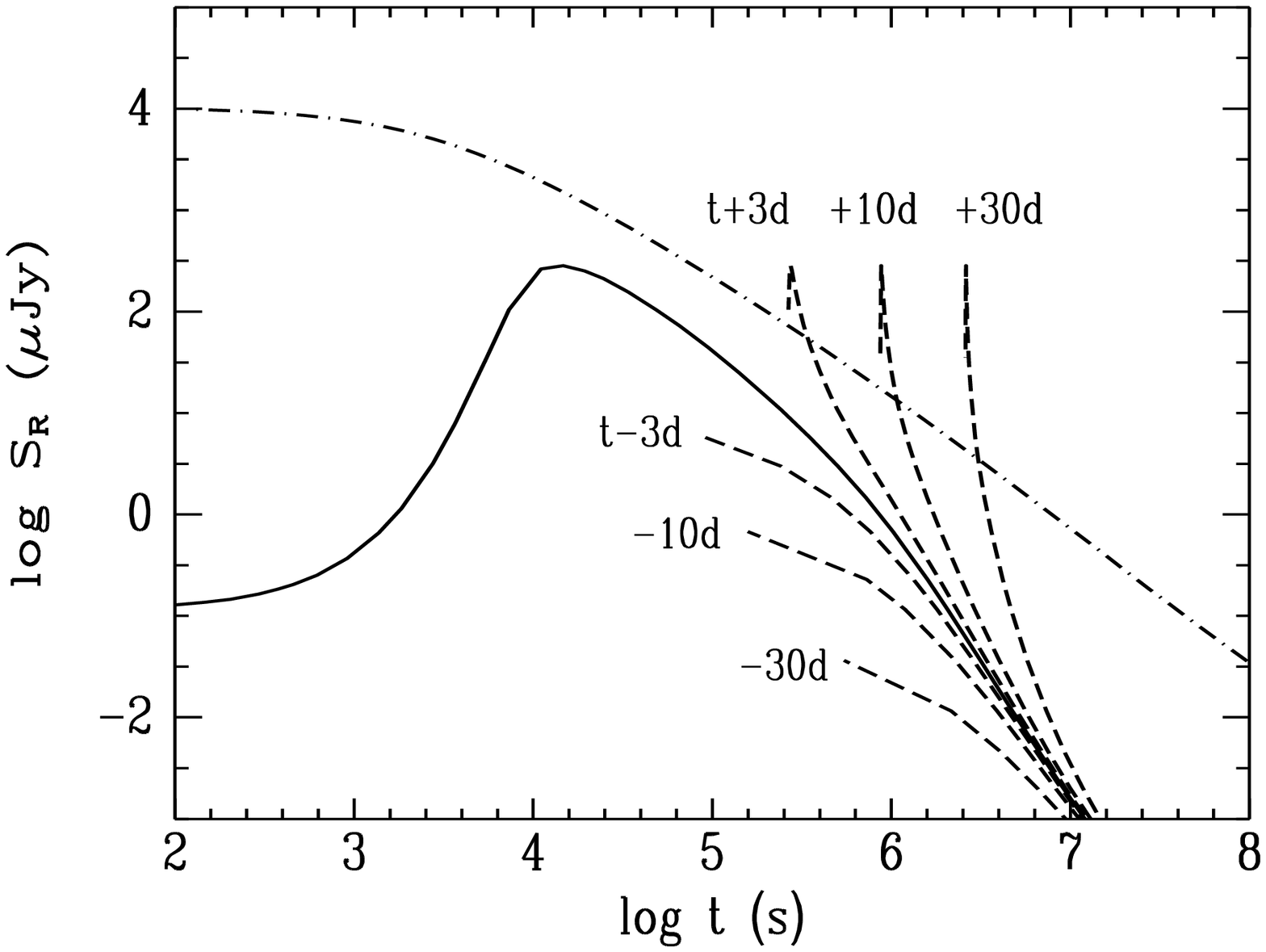}
}
  \caption{Direct comparison of the two kinds of orphan afterglows (Huang et al. 2002). 
In the left panel, a failed GRB orphan is shifted by $t \pm 3$ d, 
$t \pm 10$ d, and $t \pm 30$ d to show the effect of the uncertainty of trigger time. 
In the right panel, a jetted GRB orphan is shifted similarly. }
\end{figure}


\vspace{5mm}

\noindent
This research was supported by the National Natural Science Foundation
of China, the FANEDD (Project No: 200125), the National 973 
Project, and a RGC grant of Hong Kong SAR.


\bibliographystyle{aipproc}   

\bibliography{sample}

\section{References}

Dermer C.D., B\"ottcher M., Chiang J., 2000, ApJ, 537, 255 \\
Dermer C.D., Humi M., 2001, ApJ, 556, 479 \\
Frail D., Waxman E., Kulkarni S.R., 2000, ApJ, 537, 191 \\
Granot J., Panaitescu A., Kumar P., Woosley S.E., 2002, ApJ, 570, L61 \\
Grindlay J.E., 1999, ApJ, 510, 710 \\
Huang Y.F., 2000, astro-ph/0008177 \\
Huang Y.F., Cheng K.S., 2003, MNRAS, 341, 263 \\
Huang Y.F., Dai Z.G., Lu T., 1998, A\&A, 336, L69 \\
Huang Y.F., Dai Z.G., Lu T., 1999, MNRAS, 309, 513 \\
Huang Y.F., Dai Z.G., Lu T., 2002, MNRAS, 332, 735 \\
Huang Y.F., Gou L.J., Dai Z.G., Lu T., 2000a, ApJ, 543, 90 \\
Huang Y.F., Dai Z.G., Lu T., 2000b, MNRAS, 316, 943 \\
in't Zand J.J.M. et al., 2001, ApJ, 559, 710 \\
Lamb D.Q., 2000, Phys. Report, 333, 505 \\
Levinson A. et al., 2002, ApJ, 576, 923 \\
Livio M., Waxman E., 2000, ApJ, 538, 187 \\
M\'esz\'aros P., Rees M.J., Wijers R., 1999, New Astron., 4, 303 \\
Nakar E., Piran T., Granot J., 2002, ApJ, 579, 699 \\
Panaitescu A., Kumar P., 2003, MNRAS submitted (astro-ph/0308273) \\
Perna R., Loeb A., 1998, ApJ, 509, L85 \\
Piro L. et al., 2001, ApJ, 558, 442 \\
Rhoads J.E., 1997, ApJ, 487, L1 \\
Rhoads J.E., 2003, ApJ, 591, 1097 \\
Totani T., Panaitescu A., 2002, ApJ, 576, 120 \\
Wijers R., Rees M.J., M\'esz\'aros P., 1997, MNRAS, 288, L51 \\
Yamazaki R., Ioka K., Nakamura T., 2003, ApJ, 593, 941 \\
Zhang B., M\'esz\'aros P., 2003, Int. J. Mod. Phys. A, in press (astro-ph/0311321) 

\IfFileExists{\jobname.bbl}{}
 {\typeout{}
  \typeout{******************************************}
  \typeout{** Please run "bibtex \jobname" to optain}
  \typeout{** the bibliography and then re-run LaTeX}
  \typeout{** twice to fix the references!}
  \typeout{******************************************}
  \typeout{}
 }

\end{document}